\documentstyle[12pt,epsf]{article}
\newcommand{\beq}{\begin{equation}}
\newcommand{\eeq}{\end{equation}}

\title{ Conditional Density Matrix 
\\in  the Context of Noncontextuality}
\author{ V.Belokurov, O.Khrustalev,
 V.Sadovnichy and O.Timofeevskaya 
\\ {\em Moscow State University, 119992, Moscow, Russia}
\\ {\it e-mail: olga@goa.bog.msu.ru}}

\date{ \ \ \  }

\begin{document}
\maketitle

\begin{abstract}

 Conditional density matrix  represents 
a quantum state of subsystem in different schemes of quantum communication.
Here we discuss some properties of conditional density matrix and
its place in general scheme of quantum mechanics.
\\
\\
\\
\\
\\
\\ Talk presented at the Eleventh Lmonosov Conference on Elementary 
Particle Physics, Moscow, August, 2003.

\end{abstract}

\section{Introduction}

 A problem of a correct quantum mechanical description of
divisions of quantum systems into subsystems and reunification of
subsystems into new joint systems attracts a great interest due to
the present development
 of quantum communication.

In principle the theory of such processes was created in  general by von 
Neumann in 1927 \cite{Neu} when he constructed  deductive scheme 
of quantum  mechanics.

Nevertheless until now the description of similar processes and their
 interpretation involves some problems. For example, a recent entanglement 
swapping experiment raised discussion. The authors  \cite{Swa1} finished
their interpretation of experimental results about entanglement swapping
and teleportation by quotation: {\it this paradoxes do not arise if the
 correctness of quantum mechanics is firmly believed.(A.Peres)}. This 
statement does little to explain why there is no paradox and it is 
difficult to agree with it because quantum mechanics is a science.

Quantum mechanics predicts probabilities for various possible outcomes of
 measurement once we specify the procedure used for preparation of the 
physical system. In quantum mechanics the important assumption, which
 might be called the "noncontextuality" of probabilities, means that 
probabilities are consistent with the Hilbert-space structure of the 
observable. With these assumptions the probabilities for all measurements
can be derived from a density operator $\hat \rho$ associated by von Neumann
with quantum mechanical state.

The quantum states of subsystem of complicated system are described by reduced 
and conditional density matrices. This approach helps to avoid difficulties
and paradoxes in interpretation of some delicate experiments in different 
schemes of quantum communication.

\section{ The State in Quantum Mechanics   }

\subsection{Kinematic Quantum Mechanics Postulate}

 In quantum mechanics 

{\it each dynamical variable $\cal F$ of a system $\cal S$
corresponds to a linear operator  $\hat F$ in Hilbert space  $\cal
H$}
$$  dynamical \quad variable \quad
{\cal F} \quad \Longleftrightarrow \quad linear \quad operator
  \quad {\hat F}.
$$

To compare predictions of the theory with experimental data it was
necessary to understand how one can determine the values of
dynamical variables in the given state. W.Heisenberg \cite {Hei} gave a
partial  answer to this problem:

  {\it If   matrix that corresponds to the dynamical variable  is
diagonal, then its diagonal elements  define possible values for
the dynamical variable, i.e. its spectrum.}
$$    (\hat F)_{mn} = f_{m}{\delta}_{mn}
  \quad \Longleftrightarrow \quad
    \lbrace f_{m} \rbrace \quad is \quad spectrum \quad {\cal F}.
$$

\subsection{Quantum  State}

The general solution of the problem was given by von Neumann in
1927. He proposed the following procedure for calculation of
average values of physical variables:
$$  < {\cal F} >
    \quad = \quad
        Tr({\hat F}{\hat {\rho}}). \eqno{(1)}
$$

It is possible to represent the operator $\hat F$ in the form
$$
\hat F =\sum_{n,m} |{\psi}_n\rangle\langle {\psi}_n|\hat F
|{\psi}_m\rangle\langle {\psi}_m|=\sum_{n,m} F_{nm}\hat P_{mn},
$$
where $\{ |{\psi}_m\}$ is a basis in Hilbert space and
$$
\hat P_{mn}=|{\psi}_n\rangle\langle {\psi}_m|.$$
The average value of the variable $\hat F$ is
$$
<\hat F >=\sum_{n,m} F_{nm} {\rho}_{mn},
$$
where
$$
{\rho}_{mn}=<\hat P_{mn}>.
$$

If we suppose that the numbers ${\rho}_{mn}$ define the operator $\hat \rho :$
$ \langle {\psi}_n|\hat \rho |{\psi}_m\rangle ={\rho}_{mn}$
then the average  of the variable $\hat F$ is represented in the form (1).

Really, when they suppose that the operator $\hat \rho$ does not
depend of the variable $\hat F$ but only depends on the physical 
state of the quantum system they  introduce the proposition that
the theory is noncontextual. All subsequent experiments  confirmed
quantum mechanical  theory.

  Operator $\hat \rho$  have to satisfy three conditions:
$$   1) \quad {\hat \rho}^{+} \quad = \quad
         {\hat \rho},
$$
$$   2) \quad Tr{\hat \rho} \quad = \quad 1,
$$
$$   3) \quad  \forall \psi \in {\cal H} \quad
   <\psi|{\hat \rho}\psi>  \quad \geq 0.
$$
By the formula for average values von Neumann found out the
correspondence  between linear operators $\hat \rho$ and states of
quantum systems:
$$  \quad state \quad of\quad a\quad system \quad \rho
    \quad \Longleftrightarrow  \quad
  linear \quad operator \quad {\hat \rho}.
$$
In this way, the formula for average values becomes quantum
mechanical definition of the notion "a state of a system". The
operator $\hat \rho$ is called {\bf Density Matrix}.

If ${\hat F}$ is an observable with pure discrete spectrum
$$ {\hat F} \quad = \quad \sum_{n}f_{n}{\hat P}_{n},
$$
then
$$ \langle {\hat F} \rangle \quad  = \quad
        \sum_{n}f_{n}Tr({\hat P}_{n}{\hat \rho}).
$$
Therefore, $Tr({\hat P}_{n}{\hat \rho})$ is a probability of an observable 
 ${\hat F}$ gets a value  $f_{n}$ in the state ${\hat \rho}$.

Since density matrix is a positive definite operator and its trace
equals 1, we see that its spectrum is pure discrete  and it can be
written in the form
$$  {\hat \rho}   \quad = \quad
     \sum_{n}p_{n}{\hat P}_{n},
$$
where  ${\hat P}_{n}$ is a complete set of self-conjugate
projective operators:
$$  {{\hat P}_{n}}^{+} = {\hat P}_{n}, \quad
 {\hat P}_{m}{\hat P}_{n} = {\delta}_{mn}{\hat P}_{m},
    \quad   \sum_{n}{\hat P}_{n} = {\hat E}.
$$
Numbers $\lbrace p_{n} \rbrace$ satisfy the condition
$$  p_{n}^{*} = p_{n},  \quad 0 \le p_{n},   \quad
   \sum_{n}p_{n}\,Tr{\hat P}_{n}  = 1.
$$
It follows that $\hat \rho$ acts according to the formula
$$   {\hat \rho}{\Psi} \quad = \quad
      \sum_{n} p_{n} \sum_{\alpha \in {\Delta}_{n}}
     {\phi}_{n\alpha}\langle \phi_{n\alpha}|{\Psi} \rangle.
$$
The vectors $\phi_{n\alpha}$ form an orthonormal basis in the
space $\cal H$. Sets ${\Delta}_{n} = \lbrace 1,...,k_{n} \rbrace$
are defined by  degeneration multiplicities $k_n$ of eigenvalues
 $p_{n}$.

\subsection{Dispersion and Pure States}
From the properties of density matrix and the definition of
positively definite operators:
$$
     {\hat F}^{+} = {\hat F}, \quad \quad
    \forall \psi \in {\cal H}  \quad
      <\psi|{\hat F}{\psi}> \quad \geq 0,
$$
it follows that the average value of nonnegative variable is
nonnegative. Moreover, the average value of nonnegative variable
is equal to zero if and only if this variable equals zero.  Now it
is easy to give the following definition:

{\it  variable $\cal F$ has a
 definite value in the state $\rho$ if and only if its
 dispersion in the state  $\rho$ is equal to
 zero.
}

 The dispersion of a quantum variable $\cal F$ in the state
$\rho$ has the form:
$$   {\cal D}_{\rho}({\cal F}) \quad = \quad
         Tr({\hat Q}^{2}{\hat \rho}),
$$
where $\hat Q$ is an operator:
$$  \hat Q \quad =
\quad {\hat F} - <{\cal F}>{\hat E}.
$$
If $\cal F$ is observable ($\hat F =\hat F^+$)
 then $Q^{2}$ is a positive definite
variable. It follows that the dispersion of $\cal F$ is
nonnegative.  This makes clear the above-given definition.

 The dispersion of the observable $\cal F$ in the state $\rho$
is given by the equation
$$  {\cal D}_{\rho}({\cal F}) \quad = \quad
   \sum_{n} p_{n} \sum_{\alpha \in {\Delta}_{n}}
      ||{\hat Q}{\phi}_{n\alpha}||^{2}.
$$
All terms in this sum are nonnegative. Hence, if the dispersion is equal to
zero, then
$$  if \quad p_{n} \not= 0,  \quad
    then \quad {\hat Q}{\phi}_{n\alpha} = 0.
$$
Using the definition of the operator $\hat Q$, we obtain
$$  if \quad p_{n} \not= 0,  \quad
   then \quad {\hat F}{\phi}_{n\alpha} =
     {\phi}_{n\alpha}\langle F \rangle.
$$
In other words, {\it if an observable ${\cal F}$ has a
definite value in the given state ${\rho}$, then this value is
equal to one of the eigenvalues of the operator ${\hat F}$. }

 In this case we have
$$  {\hat \rho}{\hat
F}{\phi}_{n\alpha} \quad = \quad
    {\phi}_{n\alpha}p_{n}\langle {\cal F} \rangle\,,
\quad  {\hat F}{\hat \rho}{\phi}_{n\alpha} \quad = \quad
    {\phi}_{n\alpha}\langle {\cal F} \rangle p_{n}\,,
$$
that proves the commutativity of operators $\hat F$ and $\hat
\rho$.

It is well known, that if $\hat A$  and  $\hat B$  are commutative
self-conjugate operators, then there exists self-conjugate
operator $\hat T$ with non-degenerate spectrum such that $\hat A$
and $\hat B$ are functions of  $\hat T$.

Suppose   $\hat F$ is an operator with non-degenerate
spectrum. Then,

{\it if the observable ${\cal F}$ with non-degenerate spectrum
has a definite value in the state ${\rho}$, then it is possible to
represent the density matrix of this state as a function of the
operator ${\hat F}$. }

The operator $\hat F$ can be written in the form
$$  {\hat F} \quad = \quad
    \sum_{n}f_{n}{\hat \Pi}_{n},
$$
$$  {{\hat \Pi }_{n}}^{+} = {\hat \Pi }_{n}, \quad
 {\hat \Pi }_{m}{\hat \Pi }_{n} = {\delta}_{mn}{\hat \Pi }_{m},
     \quad tr({\hat \Pi }_{n}) = 1,
    \quad   \sum_{n}{\hat \Pi }_{n} = {\hat E}.
$$
The numbers $\lbrace f_{n} \rbrace$ satisfy the conditions
$$  f_{n}^{*} = f_{n},   \quad
   f_{n} \neq f_{n^{'}}, \quad if \quad n \neq n^{'}.
$$
From
$$  \langle F \rangle    = \sum_{n} p_{n}f_{n}    =   f_{N},
                                    \qquad
  \langle F^2 \rangle   = \sum_{n} p_{n}f_{n}^{2}=  f_{N}^{2}
$$
we get
$$   p_{n} \quad = \quad {\delta}_{nN}.
$$
In this case  density matrix is a projective operator satisfying
the condition
$$   {\hat \rho}^{2} \quad =  \quad {\hat \rho}.
$$
It acts as
$$  {\hat \rho}{\Psi} \quad =\quad
     \hat {\Pi}_N |{\Psi} \rangle =\quad
     {\Psi}_{N}\langle {\Psi}_{N}|{\Psi} \rangle,
$$
where $|{\Psi}_N \rangle$ is a vector in Hilbert space.
It is so-called {\it pure } state. 

\subsection{Density Matrix and Gleason Theorem}

 To each observable 
there corresponds a set of orthogonal projection operators $\{{\Pi}_i \} $
over a complex Hilbert space ${\cal H}$ that form a decomposition
of the identity
$$\sum_n \hat {\Pi}_n  =\hat E.$$
Quantum mechanics dictates that it is expected the various outcomes with
a probability
$$ p_n=Tr (\hat \rho \hat {\Pi}_n ).
$$
It was assumed by von Neumann and then was finally proven by Gleason\cite{Gl}
 in 1957 as the following theorem:

{\it   Assume there is a function 
$f$ from the one-dimensional projectors acting on a Hilbert space 
of dimension greater than 2 to the unit interval, with the property that 
for each orthonormal basis $\{ |{\psi}_k\rangle \} $,
$$
\sum_k f(|{\psi}_k\rangle\langle{\psi}_k)|=1.
$$
Then there exists a density matrix operator $\hat \rho$ such that}
$$
f(|{\psi}\rangle\langle {\psi}|)=\langle {\psi}|\hat \rho
|{\psi}\rangle.
$$
It assumes  that each orthonormal basis 
corresponds to mutually exclusive results of measurement of some observable.
The task is to derive the probabilities for the 
 measurement outcomes. The only  requirement is that the 
probability for obtaining the result corresponding to a normalized vector 
$|\psi\rangle$ depends only on $|\psi\rangle$ itself, not on the other 
vectors 
in the orthonormal  basis defining a particular measurement. 
This important assumption is called the "noncontextuality".
 It means that the probabilities are consistent with 
the Hilbert-space structure of observables. With these assumptions
the probabilities for all measurements can be derived from a density matrix
 using the standard quantum probability rule.

\section{Conditional Density Matrix}

\subsection{Composite System and Reduced Density Matrix}

 Suppose  that the  Hilbert space $\cal H$ is
a direct product of two  Hilbert spaces  ${\cal H}_{1}$, ${\cal H}_{2}$:
$$  {\cal H} \quad = \quad {\cal H}_{1}\otimes{\cal H}_{2}.
$$
Suppose the composite indexes  $m, \quad n, ...$  are divided  into two parts:
$m = \lbrace r, u \rbrace; n = \lbrace s, v \rbrace,...$
So, there is a basis in the space $\cal $ that can be written in
the form
$$    |\phi{\rangle}_n \quad = \quad
          |f{\rangle}_r|g{\rangle}_v .
$$
In quantum mechanics it means that the system $S$ is a unification
of two subsystems $S_{1}$  and $S_{2}$:
$$   S \quad = \quad
S_{1} \cup S_{2}\,.
$$
The Hilbert space  $\cal H$  corresponds to the system $S$ and the
spaces ${\cal H}_{1}$ and ${\cal H}_{2}$ correspond to the
subsystems $S_{1}$ and $S_{2}$.

If quantum state  of the composite
system is density matrix  $\rho_{1+2}$ then 
the state of the subsystem $S_1$ is defined by {\bf Reduced Density
matrix}
$$   {\hat \rho}_{1} \quad = \quad
      Tr_{2}{\hat \rho}_{1+2},
$$
the reduced density matrix for the subsystem $S_2$ is 
$$   {\hat \rho}_{2} \quad = \quad
      Tr_{1}{\hat \rho}_{1+2}.
$$
Quantum states  $\rho_{1}$ and $\rho_{2}$ of subsystems are
defined uniquely by the state $\rho_{1+2}$ of the composite
system.                   

\subsection{Conditional Probabilities}

We recall some definitions of probability theory.

Let $h$ be an event with positive probability. For any event A 
we define
$$
{\bf P}\{A|h\} \quad =\quad {{\bf P}\{Ah\}\over {\bf P}\{h\}}.
$$
This is conditional probability of the event A for given event $h$.

This formula can be written in the form:
$$
{\bf P}\{Ah\} \quad =\quad {\bf P}\{A|h\}{\bf P}\{h\}.
$$
Let $h_1,...h_n$ be a set of mutually exclusive events such that 
one of them takes place necessarily. Then
any event $A$ can take place only with one of the events $h_j$.
It can be written as
$$
A\quad =\quad Ah_1\cup Ah_2 \cup ...\cup Ah_n.
$$
Since $Ah_j$ are mutually independent their probabilities are added.

Thus,
$$
{\bf P}\{A\} \quad =\quad \sum_j {\bf P}\{A|h_j\}{\bf P}\{h_j\}.\eqno{(2)}
$$
This is well-known formula for total probability in terms of
conditional probabilities.

\subsection{Conditional Density Matrix}

 Let the operators  $\hat P^{(2)}_n $ be the projections on certain basis 
states in the Hilbert space
${\cal H}_2$ of pure states of subsystem  $S_2$ :
$$
\hat P^{(2)}_n\quad =\quad |u_n\rangle\langle u_n|,
\quad \sum_n \hat P^{(2)}_n =\hat E. 
$$
According definition the reduce density matrix for subsystem $S_1$ is
$$
{\rho}^{(1)}_{sr}\quad = \quad 
\sum_{uv} {\delta}_{uv} {\rho}_{sv;ru}
  \quad = \quad
\sum_{uv} \sum_n (\hat P^{(2)}_n)_{uv} (\hat \rho)_{sv;ru} =
$$
$$
\sum_n\sum_{u} (\hat P^{(2)}_n\hat \rho )_{su;ru}\quad =\quad
\sum_n p_n {\sum_{u}(\hat P^{(2)}_n\hat \rho )_{su;ru}\over p_n}.
$$
Therefore,  the reduced density matrix ${\rho}^{(1)}$
is written in the form:
$$
\hat {\rho}^{(1)}\quad = \quad 
 \sum_n p_n \hat {\rho}^{(c)}_n ,      \eqno(3)
$$
where
$$
p_n \quad =\quad 
 \sum_{uv}P^{(2)}_{n}(u|v)\sum_{r}{\rho}_{rv;ru}
      \quad = \quad
       \sum_{uv}P^{(2)}_{n}(u|v){\hat \rho}^{(2)}(v,u)
$$
or
$$
p_n\quad =\quad Tr_2 (\hat P^{(2)}_n{\hat \rho}^{(2)}).
$$
If the set of projections ${\hat P^{(2)}}_{n}$ is associated with
some observable in the subsystem $S_{2}$ 
$$   {\hat G} \quad = \quad \sum_{n}g_{n}{\hat P^{(2)}}_{n}, 
$$
then $p_{n}$ is a probability of the variable ${\hat G}$ gets a value
$g_{n}$ in the state ${\hat \rho}^{(2)}$.

The operator $\hat {\rho}^{(c)}_{n} $ equals:
$$  {\rho}^{(c)}_{n}(r|s) \quad = \quad {1 \over p_{n}} 
    \sum_{uv}P^{(2)}_{n}(u|v){\rho}_{rv;su}
$$
and  satisfies all conditions (1). It is {\bf density matrix} or
quantum state.

Since $w_{r}={\hat \rho}^{(1)}_{rr}$ is a probability to find a subsystem
$S_1$ in the state $|r \rangle$,
we see that an equality
$$  w_{r} \quad = \quad 
    \sum_{n}p_{n}{\rho}^{(c)}_{n}(r|r), 
$$    
is formula(2). 

Then the operator ${\hat \rho}^{(c)}_{n}$ is called {\bf conditional
 density matrix} and is written  \cite {Sol}
$$
{\hat \rho}^{(c)}_{1/2n}\quad =\quad
{Tr_2 (\hat P^{(2)}_n \hat \rho) \over Tr(\hat P^{(2)}_n \hat
\rho ) }\quad=\quad
{Tr_2 (\hat P^{(2)}_n \hat \rho) \over w_n }.      \eqno{(4)}
$$
This is a {\bf conditional density matrix}, i.e. {\bf a quantum state},
for subsystem $S_1$ under condition that the subsystem $S_2$ is selected
in {\bf pure state} $\hat P^{(2)}_n$. 
It is the most interesting case for quantum communication.
 This definition of the quantum state of
quantum subsystem assumes   noncontextual approach in quantum mechanics.

It is necessary to note that although  formula (4)  arose in description 
of measurement, for example in papers \cite {Bar},
\cite {Fuchs} and et. ,
it was presented as the result 
of transformation of quantum state of the system during measurement. 
Here, conditional density matrix is the definition of a new quantum state of
 the subsystem that is selected under definite physical condition.

\subsection{Conditional Density Matrix in Case of Generalized
Measurement}

It was recently shown \cite{Busch} that a Glison-like theorem can be easy
proved (and also extends to the case of 2-dimensional Hilbert space) 
on a set of {\it effects} $\{ E\}$.  It is a set of projections 
but commutativity (or orthogonality) is no longer necessary. 
According to this theorem

 {\it any generalized probability
 measure is of the form $E\rightarrow v(E)=tr[\rho E]$ for all $E$,
 for some density operator $\rho .$}

\noindent While we consider a generalized measurement \cite{Fuchs}  in the subsystem
$S_2$ we suppose that a set of projections $\{\hat E_b\}$ exits and 
 satisfies the properties
$$
<\psi |\hat E_b |\psi >\quad \ge \quad 0,\quad 
\forall |\psi >,\quad \sum_b \hat E_b =\hat I^{(2)} .
$$
The probabilities of outcomes are equal 
$$
P(b)=tr(\hat \rho\hat E_b ).
$$
We don't suppose that condition ${\Pi}_i{\Pi}_j={\delta}_{ij}{\Pi}_i$
is fulfilled.

In this case the decomposition (3) is also valid and quantum state
of subsystem $S_1$ under condition that the subsystem  $S_2$ is selected
in pure state $\hat E_b$ is
$$
{\hat \rho}^{(c)}_{1/2b}\quad =\quad
{Tr_2 (\hat E_b \hat \rho) \over Tr(\hat E_b \hat
\rho ) }.
$$

\section{Conditional Density Matrix Description  of Entanglement
Swapping }

In the experiments  \cite{Swa1} with installation  two pairs of
correlated photons are emerged simultaneously. The polarization state of the
system is being described by the simultaneous wave function
$$   
   |{\Psi}({\sigma}_{1}, {\sigma}_{2}, {\sigma}_{3}, {\sigma}_{4}) \rangle
            \quad =
\quad {\Psi}_-({\sigma}_{1},{\sigma}_{2}){\Psi}_-({\sigma}_{3},{\sigma}_{4}),
$$
where ${\Psi}_-$ is antisymmetric state of pair of photons 
$$
{\Psi}_-({\sigma}_i,{\sigma}_j)=
={1\over \sqrt 2}\big( {\chi}_0 ({\sigma}_i){\chi}_1 ({\sigma}_j)-
{\chi}_1 ({\sigma}_i){\chi}_0 ({\sigma}_j)\big)
$$
and ${\chi}_s ({\sigma})$ are two basis states with orthonormal polarization.
Reduced density matrix of subsystem $S_{1-4}$ is proportional to unity
$$ {\rho}_{14}={1\over 2}\hat I^{(1)} \otimes {1\over 2}\hat I^{(4)}.
$$

But if we select the pair of photons 1-4 only under condition that
pair 2-3 is in the pure state ${\Psi}_-({\sigma}_2,{\sigma}_3)$ then 
quantum state of pair 1-4 is conditional density matrix
$$  {\hat \rho}^c_{14/23} \quad = \quad
  {Tr_{23}({\hat P}_{23}{\hat \rho}_{1234}) \over
  Tr({\hat P}_{23}{\hat \rho}_{1234})},
$$
where operator ${\hat P}_{23}$ selects pair 2-3 is pure state 
${\Psi}_-({\sigma}_2,{\sigma}_3)$.
Direct calculation shows that the pair of the photons (1 and 4)
has to be in pure state with the wave function
$
\Phi({\sigma}_{1},{\sigma}_{4}) \quad = \quad
      {\Psi}_-({\sigma}_{1},{\sigma}_{4}).
$
As the system $S_{1234}$ is described by simultaneous wave function 
the time order of measurements has no importance.

Other examples demonstrating the utilization of conditional density matrix
in different schemes of quantum communication are represented in \cite {Ech}.

\section{Conclusion}

 Provided that the  subsystem $S_2$ of composite quantum system $S=S_1 + S_2$
is selected in a pure state $\hat P_n$
the quantum state of subsystem
$S_1$ is conditional density matrix  $\hat {\rho}_{1c/2n}$. Reduced
density matrix $\hat {\rho }_1$ is connected with conditional
density matrices by an  expansion (3).

\end{document}